# Self-referencing a continuous-wave laser with electro-optic modulation

Katja Beha, Daniel C. Cole, Pascal Del'Haye,
Aurélien Coillet, Scott A. Diddams, and Scott B. Papp*
Time and Frequency Division, National Institute of Standards and Technology, 325 Broadway,
Boulder, CO 80305 USA
*scott.papp@nist.gov

We phase-coherently measure the frequency of continuous-wave (CW) laser light by use of optical-phase modulation and 'f-2f' nonlinear interferometry. Periodic electro-optic modulation (EOM) transforms the CW laser into a continuous train of picosecond optical pulses. Subsequent nonlinear-fiber broadening of this EOM frequency comb produces a supercontinuum with 160 THz of bandwidth. A critical intermediate step is optical filtering of the EOM comb to reduce electronic-noise-induced decoherence of the supercontinuum. Applying f-2f self-referencing with the supercontinuum yields the carrier-envelope offset frequency of the EOM comb, which is precisely the difference of the CW laser frequency and an exact integer multiple of the EOM pulse repetition rate. Here we demonstrate absolute optical frequency metrology and synthesis applications of the self-referenced CW laser with <5 x $10^{-14}$ fractional accuracy and stability.

Self-referencing an optical frequency comb by f-2f nonlinear interferometry provides a cycle-by-cycle link between lightwave and microwave frequencies (1, 2). Such devices have enabled myriad applications from optical clocks (3) to precisely calibrated spectroscopy (4–7) to photonic microwave generation (8). Moreover experimental control of carrier-envelope offset phase contributes to ultrafast science (9). Self-referenced combs are produced with various modelocked laser platforms by leveraging spectral broadening with typically sub-GHz repetition rates and <100 fs duration optical pulses (10, 11). Still, realizing new applications calls for the development of easy-to-operate frequency combs that offer a widely tunable spectrum and large mode frequency spacing (12, 13).

Optical frequency combs are alternatively realized by direct electro-optic modulation of continuous-wave light (14–18). The frequency of each EOM-comb line ($N$) is $\nu_N = \nu_p + N f_{eo}$ and arises jointly from the CW-laser frequency $\nu_p$, and from frequency multiplication of the modulation $f_{eo}$. Importantly, the line-by-line spectral phase of such EOM combs is predetermined by the modulation format apart from any modelocking mechanism. These frequency combs are already a driver of applications such as optical arbitrary waveform control (19), coherent communications (20), and spectroscopic measurements that require 10's of GHz comb-mode spacing and wide tuning range for calibration (21). Moreover, they can be realized with commercial components, and straightforwardly reconfigured. Despite these advantages EOM-comb offset-frequency detection via self-referencing, and hence optical measurements and synthesis with an EOM comb, has been an outstanding challenge.

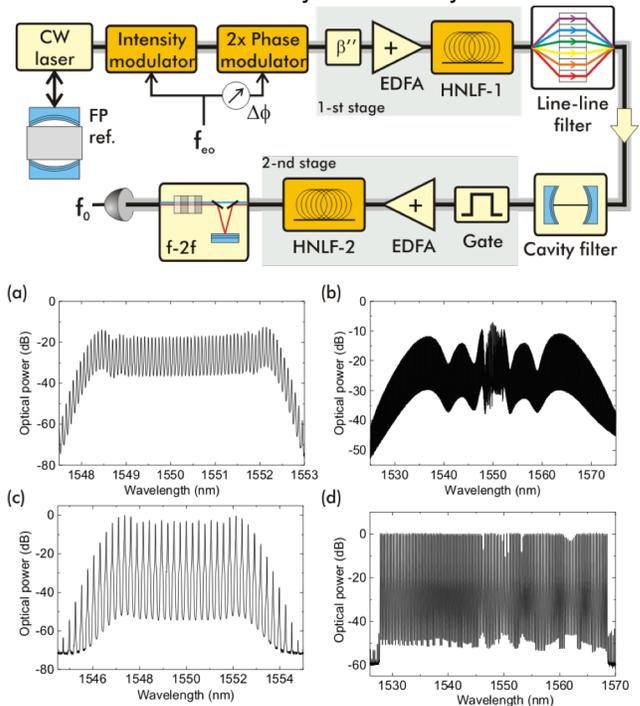

Figure 1: Apparatus for self-referencing a CW laser. The system is comprised of electro-optic phase an intensity modulators, two stages of nonlinear-fiber (HNLF-1 and -2) pulse broadening, a Fabry-Perot optical cavity filter, and f-2f detection. (a) EOM comb with 10 GHz spacing. (b) Idem after HNLF-1. (c) EOM comb with 33 GHz spacing. (d) Idem after HNLF-1 and line-line intensity control.

Here we accomplish self-referencing of an EOM comb by identifying and solving two major challenges: (1) Low spectral-broadening efficiency associated with limited EOM-comb bandwidth and microwave repetition rates; and (2) Conversion of electronic noise to the optical comb that progressively degrades as $N^2$ the first-order optical coherence of the comb lines. Our approach utilizes two stages of nonlinear-fiber broadening, and an optical cavity to filter high-frequency electro-optic noise from the comb. The carrier-envelope offset frequency of an EOM comb is $f_0 = \nu_p - N f_{eo}$, whereby a very large multiplicative factor $N$ (~19 340 or 86 dB for our experiments with 10 GHz spacing) relates the phase-noise spectrum of the electronic oscillator to that of $f_0$. The need for optical filtering arises since microwave oscillators, given their mostly constant thermal phase noise, do not support frequency multiplication to the optical domain (20, 22). Optical filtering reduces the impact of high-frequency oscillator noise and modifies the phase noise Fourier-frequency dependence to $1/f^2$, which is consistent with that of a laser.

One method to understand the filter cavity's effect is to note that solution of an integral equation (23) $\int_{\delta\omega}^{2\pi f_{eo}/2} S_\phi^{(N)} F(\omega)\, d\omega = 1$ estimates the linewidth $\delta\omega^{(N)}$ of the $N$-th EOM comb line. Here we focus on the limiting case for an EOM-comb mode with constant phase noise $S_\phi^{(N)} = N^2 S_\phi$, where $S_\phi$ is a constant oscillator spectrum and $F(\omega)$ is the filter cavity lineshape that we approximate as a Heaviside function. Even for a state-of-the-art 10 GHz oscillator with constant thermal-limited phase noise of -188 dBc/Hz at +12 dBm, without filtering, the comb lines needed for self-referencing are estimated to have $\delta\omega^{(19\,340)}/2\pi > 1$ GHz and are practically undetectable. By use of a ~10 MHz bandwidth optical cavity following EOM-comb generation, we significantly reduce the contribution from constant oscillator phase noise. The result is a negligible projected linewidth contribution to the EOM-comb modes needed for self-referencing.

Our EOM-comb system (Fig. 1) begins with a 1550 nm CW laser that is for convenience stabilized to a high-finesse, low-expansion Fabry-Perot cavity (24). A frequency comb is derived by way of optical-phase modulation with waveguide lithium-niobate devices at frequency $f_{eo}$ that transforms the CW laser into light pulses repeated with each modulation cycle. The modulation is derived from a hydrogen-maser referenced commercial frequency synthesizer. Intensity modulation carves out of the CW laser a train of 50 % duty cycle pulses with >99 % contrast. This pulse train passes through two sequential phase modulators, which impart a chirp to each optical pulse. The relative modulator phases of are chosen such that the carrier frequency of the pulses increases linearly in time when the pulse intensity is highest. This linear chirp yields pulses whose spectral envelopes are relatively flat, as shown in Figs. 1a and 1c for combs with 10 GHz and 33 GHz spacing, respectively. Still, the bandwidth of these combs is <1 THz, a factor of 200 too small for self-referencing. Subsequent spectral broadening of the EOM comb exploits the fact that their spectral-phase profile can be compensated using only second-order dispersion (25). Propagation of the 50 % duty cycle pulse train through an appropriate length of 1550 nm single-mode fiber (SMF) allows pulse compression to near the Fourier-transform limit.

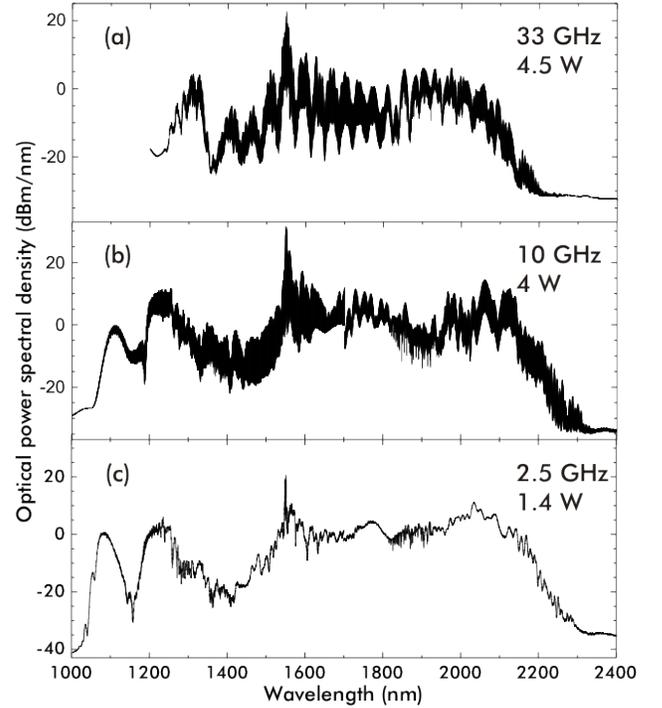

Figure 2: Supercontinuum generated using our second broadening stage with HNLF-2. (a) EOM-comb spacing is 33 GHz. (b) Idem but for 10 GHz spacing. (c) Pulse picking prior to HNLF-2 reduces the initial spacing to 2.5 GHz.

To increase the EOM-comb bandwidth for self-referencing, we utilize two stages of spectral broadening (26) in commercially available highly nonlinear fiber (HNLF). Our work opens a new regime in which we realize coherent, octave-spanning frequency combs seeded by relatively long optical pulses – 1.5 ps in the case of our 10 GHz system. In the first broadening stage, we amplify the EOM comb to 500 mW using a commercial erbium-doped fiber amplifier (EDFA). Next, the light is guided through a 100 m length of near-zero-dispersion HNLF. The resulting optical spectrum is characteristic of self-phase modulation (SPM) (27); Fig. 1b shows the 10 GHz EOM comb following the first broadening stage.

Some comb applications are not compatible with the ~10 dB power variation caused by spectral interference of temporally separated pulse components that receive similar SPM chirp. Line-by-line pulse shaping (*28*) with a liquid crystal device can correct the power variation; e.g. Fig. 1d shows the SPM spectrum of a 33 GHz EOM comb after line-by-line power flattening. Furthermore, line-by-line phase control enables us to temporally compress the EOM combs after SPM. In particular, for both 10 and 33 GHz EOM combs we can reduce the FWHM optical intensity autocorrelation to below 300 fs by applying a second-order dispersion correction. For all the experiments described below, we use the pulse shaper only to correct second-order dispersion, and the line-line controller could most likely be replaced with an appropriate length of SMF.

Our second broadening stage is designed to coherently increase the EOM-comb span to the >1000 nm needed for f-2f self-referencing. Operationally, as shown in Fig. 1, the EOM comb light is re-amplified with a cladding-pumped, anomalous dispersion Er/Yb co-doped fiber to attain 140 pJ, 400 pJ, and 560 pJ pulses for the 33 GHz, 10 GHz, 2.5 GHz EOM comb repetition frequencies, respectively. The 2.5 GHz comb is obtained by attenuating three out of every four 10 GHz pulses using a waveguide lithium-niobate intensity modulator (*9, 16, 29, 30*). The Er/Yb amplifier offers a maximum average power of 4.5 W, while the temporal intensity autocorrelation of optical pulses exiting the amplifier feature <300 fs duration. Prior to the second-stage amplifier, we adjust both the polarization and the dispersion of the second stage to maximize supercontinuum generation in HNLF-2. Specifically, second-order dispersion is applied in increments of 0.005 $ps^2$ using the line-line pulse shaper.

The design and implementation of coherent supercontinuum generation at high repetition rate is an advance of our work. We apparently avoid supercontinuum decoherence mechanisms typically associated with >100 fs pulse duration (*31*). Figure 2 shows particular examples of the ultrabroad spectra we obtain with our EOM comb. The 10 GHz and 2.5 GHz cases feature more than one octave of bandwidth, and the 10 GHz and 33 GHz comb teeth are prominent across the entire spectrum. (Optical spectrum analyzer resolution prevents observation of the 2.5 GHz spaced teeth.) HNLF-2 is composed of two segments of nonlinear fiber with different spectral-dispersion profiles. These are fusion spliced together with >80% transmission. The first segment has a dispersion zero near 1300 nm that explains the dispersive wave centered at 1100 nm in Figs 2b and

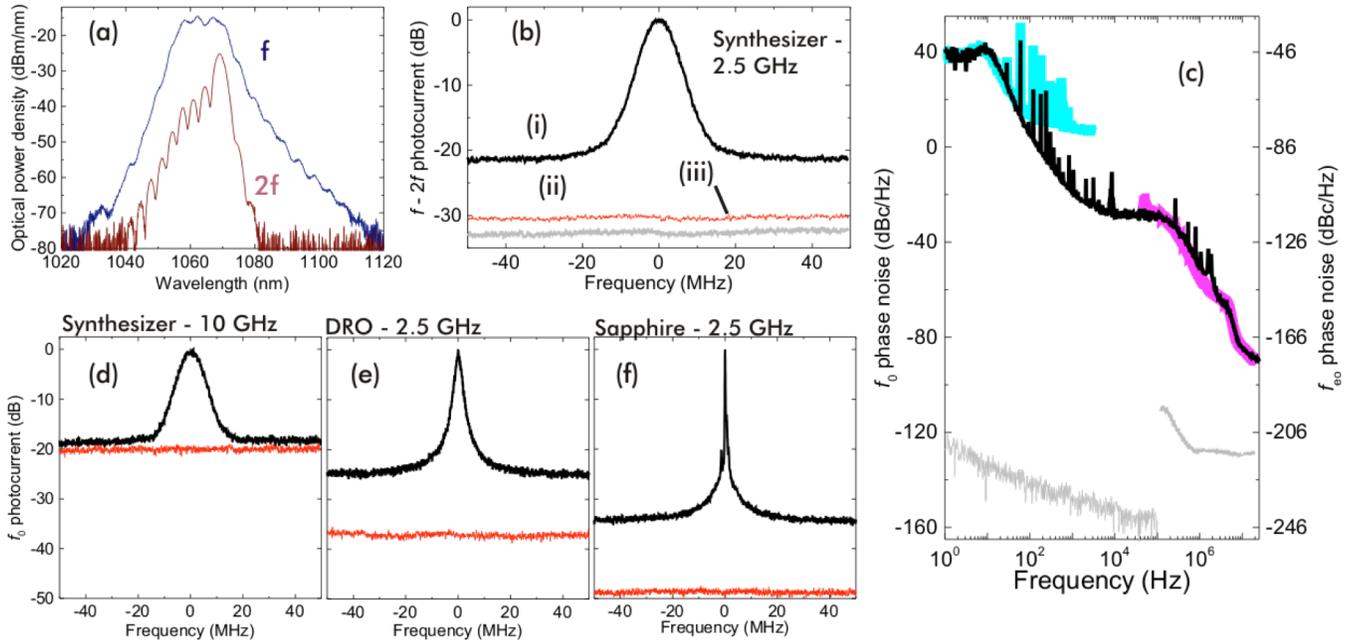

Figure 3: EOM-comb offset-frequency detection. (a) Spectra of fundamental and second harmonic supercontinuum light at 1070 nm. (b) i: Carrier-offset heterodyne beat ($f_0$) signal, ii: intensity noise, iii: detector noise. (c) Single-sideband phase noise of $f_0$. Upper cyan trace is the maser contribution and lower magenta trace is the $f_{eo}$ generator including the optical filter cavity. The grey traces indicate the phase noise detection floor, which is effectively below -200 dBc/Hz for $f_{eo}$. (d,e,f) Spectrum of $f_0$ with different $f_{eo}$ sources, including a synthesizer, a dielectric-resonator oscillator, and a sapphire-cavity oscillator. The 2.5 GHz cases utilize pulse picking. All the photocurrent spectra are acquired with 100 kHz resolution bandwidth.

c. The dispersion of the second segment is 1.5 ps/nm-km, and this fiber generates the dispersive-wave pattern near 1200 nm and the broad supercontinuum peaks near 2100 nm.

Following supercontinuum generation, we detect the carrier-envelope-offset frequency of the EOM comb. We deliver the supercontinuum from HNLF-2 to a standard free-space f-2f setup; see Fig. 1. A 10 mm sample of periodically poled lithium niobate generates the second harmonic of supercontinuum light at 2140 nm; Fig 3a shows separately obtained optical spectra of the f and 2f components at 1070 nm for the 2.5 GHz comb. Both the f and 2f laser beams pass through an optical interference filter and are re-coupled into SMF. We photodetect the optical heterodyne beat of these two spectra and optimize alignment of their relative arrival time and polarization. The photocurrent reveals for the first time the offset frequency $f_0$ of an EOM comb; Fig. 3b shows the RF spectrum of the offset beat.

We have not been able to detect $f_0$ with the filter cavity removed from the system. This Fabry-Perot optical cavity features a 7 MHz FWHM linewidth and its insertion loss is 8 dB for the entire SPM-broadened EOM comb. Our unsuccessful search for $f_0$ with the filter cavity removed involves careful re-optimization of the pulse shaper dispersion setting and all other relevant system parameters. The procedure yields f and 2f supercontinuum components comparable to those shown in Fig. 3a. (Furthermore, we use the highest performance $f_{eo}$ oscillator available; see discussion of the sapphire device below.) This result matches our expectation that the coherence of EOM-comb lines needed for self-referencing is severely degraded by frequency multiplication of broadband electro-optic comb noise.

We characterize $f_0$ in detail to understand the potential for precision optical measurements with an EOM comb. Electronic spectrum analyzer traces of the 2.5 GHz comb $f_0$ beat and its background contributions are shown in Fig. 3b; the black trace (i) is the signal, and the red (ii) and gray (iii) traces represent the supercontinuum intensity noise and photodetector noise, respectively. (Fig. 3d shows the 10 GHz case, without pulse picking.) Since the EOM comb spacing $f_{eo} = 9.999\,952$ GHz is locked to a hydrogen-maser frequency reference traceable to the International System (SI) second, detecting the center of the $f_0$ spectrum represents an absolute determination of $\nu_p$. Frequency counting experiments with $f_0$ are enabled by a high signal-noise ratio (SNR) and a narrow spectral width. The noise floor of $f_0$ exceeds the intensity noise by 8 dB. At present we cannot determine the relative contributions here from supercontinuum generation, which often induces phase noise in $f_0$ (*31*, *32*), and electronic noise that is not sufficiently attenuated by the filter cavity. Further the 7 MHz linewidth of $f_0$ exceeds that observed for many modelocked lasers. Still, as we show below, the offset beat we obtain is sufficient to establish a phase-coherent link between the microwave and optical domains.

To understand the $f_0$ lineshape, we measure its phase-noise spectrum for Fourier frequencies from 1 Hz to 20 MHz; see the black trace in Fig. 3c. Here, the phase noise of this effective 193 THz signal is overwhelmingly due to $f_{eo}$ once frequency multiplication by 19 340 is included. Specifically, the cyan trace is the hydrogen-maser contribution to $f_{eo}$ that has been adjusted by +146 dB for comparison with the black trace, and the magenta trace is the high-offset phase noise of $f_{eo}$ adjusted by +86 dB and filtered by the 7 MHz FWHM Lorentzian lineshape of the optical-filter cavity. Conversely, the phase-noise contribution from the cavity-stabilized CW laser is not visible. Agreement between the measured phase noise of $f_0$ and its known contributions is an indicator of the phase-coherent link between microwave and optical domains provided by the EOM comb. Moreover, recording the phase-noise spectrum of $f_{eo}$ in this way offers an extraordinary detection limit below -200 dBc/Hz at microwave frequencies; the phase-noise floors of our two analyzers are shown by the gray traces in Fig. 3c. This represents a key feature of EOM combs and other combs generated from an ultrastable CW laser (e.g. Kerr microcombs) that are fundamentally based on optical modulation and frequency multiplication of the comb line spacing.

The linewidth of $f_0$ is an important consideration for future applications. Therefore we study the connection between $f_0$ linewidth and its phase-noise spectrum. Previous work has analyzed the relationship between a laser's linewidth and its phase-noise spectrum (*33–35*). Building on Ref. (*35*), we would expect that the EOM comb $f_0$ linewidth is largely determined by the level of $f_{eo}$ phase noise at fourier frequencies from approximately 10 kHz to 10 MHz. (The filter cavity reduces >3.5 MHz noise.) In this range, a high modulation index leads to the most significant $f_0$ broadening. To directly explore this concept with the EOM comb, we record the power spectrum of $f_0$ using three separate $f_{eo}$ oscillators, which have different 10 kHz to 10 MHz phase-noise levels. (They are all phase locked close-to-carrier to the maser reference.) Figures 3d, 3e, and 3f are acquired using $f_{eo}$ oscillators with improving phase-noise performance: the same synthesizer as above but excluding pulse picking; a 10 GHz dielectric-resonator oscillator (DRO); and a 10 GHz sapphire-cavity oscillator (sapphire), respectively.

The data traces in Fig. 3d-f indicate up to a factor of 40 reduction in linewidth to ~100 kHz with improved $f_{eo}$ performance, which roughly corresponds to the specification of these devices (*36*). In particular, since pulse picking only divides the pulse-repetition rate, we expect and observe no linewidth change when using the synthesizer for the 10 GHz and 2.5 GHz combs. We were not able to make definitive phase-noise measurements of the DRO and sapphire oscillators in the microwave domain. Still, our $f_0$ linewidth measurements suggest the DRO (sapphire) devices feature a 10 kHz to 10 MHz integrated phase noise at least a factor of ~9 (~2200) lower than the synthesizer. Moreover these data were all acquired by use of the filter cavity; hence they highlight the role of frequency dependence $f_{eo}$ phase noise apart from broadband oscillator noise. Future experiments utilizing ultralow-noise microwave references could realize a <100 kHz $f_0$ beat that is competitive with modelocked lasers (*37*, *38*).

Access to the offset frequency of the EOM comb enables precision experiments, including measurement and synthesis of optical frequencies, which are presented here. The diagram in Fig. 4a summarizes the optical and electronic connections for our experiments with the synthesizer-driven comb. By frequency counting $f_0$ we measure $v_p$ with respect to the EOM-comb spacing. A frequency counter records $f_0$ after RF filtering and digital frequency division, which phase-coherently reduce fluctuations. Each 1 s measurement yields $v_p$ with a fractional uncertainty of $3 \times 10^{-13}$. As a crosscheck and for comparison we obtain a separate measurement of $v_p$ with respect to the 250.32413 MHz spacing of a self-referenced erbium-fiber frequency comb. Both the EOM comb and the fiber comb spacings are referenced to the same hydrogen maser. Figure 4b shows a record over 4000 s, during which we monitor the CW laser frequency with the EOM comb (black points) and fiber comb (green points). This data tracks the approximately 65 mHz/s instantaneous linear drift rate of the cavity-stabilized $v_p$. Moreover, linear fitting of the two data sets reveals an average offset between them of 17 Hz, which is within 1.8 standard deviations of the mean of their combined uncertainty.

To understand the uncertainty of such optical frequency-counting measurements, we present their Allan deviation in Fig. 4c. For less than 100 s averaging, both the EOM and fiber comb instabilities are consistent with their common reference maser's instability of $3 \times 10^{-13}/\sqrt{\tau}$, where $\tau$ is the averaging time. Beyond $\tau \sim 100$ sec the drift rate of $\nu_p$, indicated by the solid line, becomes apparent. By post-correcting the two datasets with the fitted drift rates, the fractional instability reduces to $3 \times 10^{-14}$ at 1000 s and represents the uncertainty in our determination of the means of the two sets.

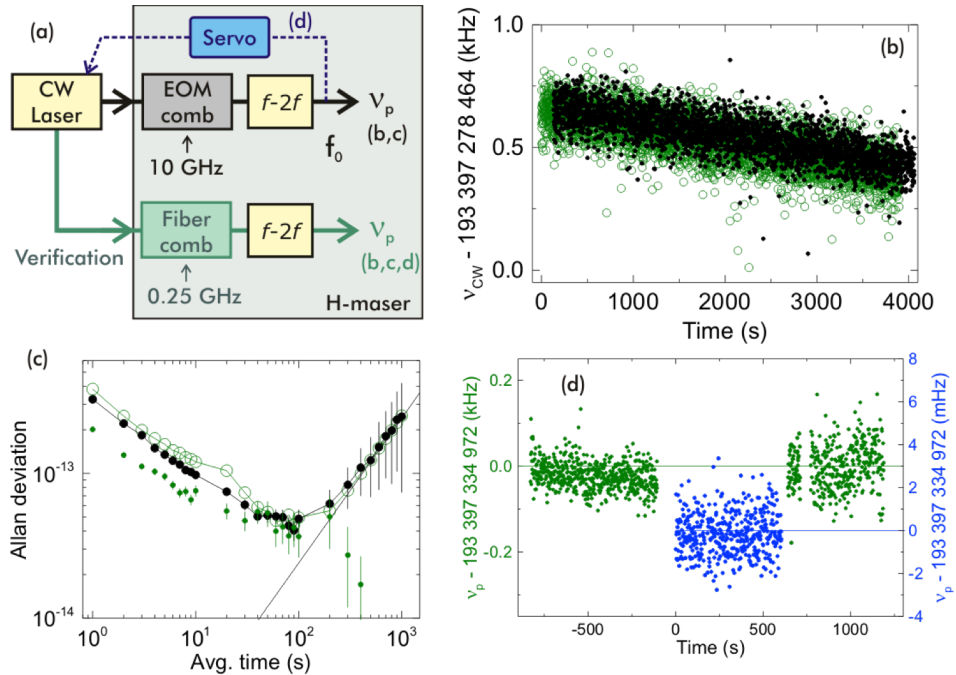

Figure 4: Demonstration of optical frequency measurements and synthesis. (a) System components and connections. Both the EOM comb and an auxiliary fiber comb are referenced to a hydrogen maser. (b) Frequency counting record over 4000 s of $v_p$ with the EOM (black points) and fiber (green circles) combs. (c) Allan deviation of the frequency count record from (b) and the synthesis data in (d). The solid line indicates the Allan deviation of a 193 THz laser drifting at 65 mHz/s. (d) Phase-lock synthesis of the CW laser frequency with respect to the maser-referenced $f_{eo}$. Green points are measurements of locked $v_p$ with the fiber comb, and blue points are the in-loop noise.

Finally, we further stabilize the frequency $\nu_p$ of the CW laser using an RF phase-lock of $f_0$ while $f_{eo}$ is held fixed at 9.999 952 GHz. Operationally, we phase-coherently filter $f_0$ and divide it by 512 prior to phase discrimination, and feedback is provided to an acousto-optic frequency modulator immediately following the CW laser. To verify the phase lock, Figure 4d presents two consecutively obtained frequency-counting modalities with $\nu_p$: fiber comb measurements (green points) and in-loop fluctuations in the $\nu_p$ phase lock (blue points). In these experiments the set point of the $\nu_p$ phase lock is held constant at precisely 193.397 334 972 THz, but this value is adjustable within the CW laser's tuning range. The data is semi-continuously acquired at 1 s gate time for 2000 s as we reconfigure system connections and slightly adjust the phase-lock loop filter. The mean difference of the green points and the phase-lock setpoint is -14(8) Hz, while the frequency offset of the in-loop signal scatters closely about zero. Furthermore, the Allan deviation of the locked $\nu_p$ is presented alongside previously described data in Fig 4c. Here the solid green points fall slightly below the other two Allan deviation measurements and do not show the 65 mHz/s cavity drift, since all system components are phase-locked to the same maser reference. This demonstrates the capability to directly synthesize with respect to the maser the optical frequency of every EOM comb line.

In conclusion, we have demonstrated self-referencing of a CW laser by way of electro-optic modulation. We form an EOM comb initially with ~600 GHz bandwidth, then use HNLF-based pulse broadening to create an octave-spanning supercontinuum. A second advance of our work is the demonstration of optical filtering to preserve the optical coherence of the EOM comb following frequency multiplication from 10 GHz to 193 THz. Using the self-referenced EOM comb, we demonstrate both precision optical frequency measurements and phase-lock synthesis of all the comb frequencies. These capabilities have already matured in the modelocked laser platform (*12*). Still, EOM combs offer a number of important advantages, including wide microwave-rate mode spacing, frequency tuning of the comb, and already commercialized systems. In addition, our work opens the capability for ultraprecision microwave phase-noise metrology by leveraging massive frequency multiplication for access to levels below -200 dBc/Hz on a 10 GHz source.

We thank L. Sinclair and E. Lamb for helpful comments on the manuscript, M. Hirano for providing the HNLF, and A. Hati for providing the sapphire oscillator. This work is supported by the DARPA QuASAR program, AFOSR, NASA, and NIST. DC acknowledges support from the NSF GRFP under Grant No. DGE 1144083. It is a contribution of the US government and is not subject to copyright in the United States.

**References**
1. S. A. Diddams *et al.*, Direct Link between Microwave and Optical Frequencies with a 300 THz Femtosecond Laser Comb. *Phys Rev Lett.* **84**, 5102–– (2000).

2. D. J. Jones *et al.*, Carrier-Envelope Phase Control of Femtosecond Mode-Locked Lasers and Direct Optical Frequency Synthesis. *Science.* **288**, 635–639 (2000).

3. S. A. Diddams *et al.*, An Optical Clock Based on a Single Trapped 199Hg+ Ion. *Science.* **293**, 825–828 (2001).

4. M. J. Thorpe, K. D. Moll, R. J. Jones, B. Safdi, J. Ye, Broadband Cavity Ringdown Spectroscopy for Sensitive and Rapid Molecular Detection. *Science.* **311**, 1595–1599 (2006).

5. M. J. Thorpe, D. Balslev-Clausen, M. S. Kirchner, J. Ye, Cavity-enhanced optical frequency comb spectroscopy: application to human breath analysis. *Opt. Express.* **16**, 2387 (2008).

6. S. A. Diddams, L. Hollberg, V. Mbele, Molecular fingerprinting with the resolved modes of a femtosecond laser frequency comb. *Nature.* **445**, 627–630 (2007).

7. G. B. Rieker *et al.*, Frequency-comb-based remote sensing of greenhouse gases over kilometer air paths. *Optica.* **1**, 290 (2014).

8. T. M. Fortier *et al.*, Generation of ultrastable microwaves via optical frequency division. *Nat*

*Photon.* **5**, 425–429 (2011).

9. A. Baltuska *et al.*, Attosecond control of electronic processes by intense light fields. *Nature.* **421**, 611–615 (2003).

10. L. C. Sinclair *et al.*, Operation of an optically coherent frequency comb outside the metrology lab. *Opt. Express.* **22**, 6996 (2014).

11. M. Akbulut *et al.*, Measurement of carrier envelope offset frequency for a 10 GHz etalon-stabilized semiconductor optical frequency comb. *Opt. Express.* **19**, 16851 (2011).

12. S. A. Diddams, The evolving optical frequency comb. *J Opt Soc Am B.* **27**, B51 (2010).

13. N. R. Newbury, Searching for applications with a fine-tooth comb. *Nat. Photonics.* **5**, 186–188 (2011).

14. M. Kourogi, K. Nakagawa, M. Ohtsu, Wide-span optical frequency comb generator for accurate optical frequency difference measurement. *Quantum Electron. IEEE J. Of.* **29**, 2693–2701 (1993).

15. I. Morohashi *et al.*, Widely repetition-tunable 200 fs pulse source using a Mach-Zehnder-modulator-based flat comb generator and dispersion-flattened dispersion-decreasing fiber. *Opt. Lett.* **33**, 1192–1194 (2008).

16. A. Ishizawa *et al.*, Octave-spanning frequency comb generated by 250 fs pulse train emitted from 25 GHz externally phase-modulated laser diode for carrier-envelope-offset-locking. *Electron. Lett.* **46**, 1343 (2010).

17. V. R. Supradeepa, A. M. Weiner, Bandwidth scaling and spectral flatness enhancement of optical frequency combs from phase-modulated continuous-wave lasers using cascaded four-wave mixing. *Opt. Lett.* **37**, 3066–3068 (2012).

18. T. Sakamoto, T. Kawanishi, M. Izutsu, Asymptotic formalism for ultraflat optical frequency comb generation using a Mach-Zehnder modulator. *Opt. Lett.* **32**, 1515–1517 (2007).

19. S. T. Cundiff, A. M. Weiner, Optical arbitrary waveform generation. *Nat. Photonics.* **4**, 760–766 (2010).

20. A. Ishizawa *et al.*, Phase-noise characteristics of a 25-GHz-spaced optical frequency comb based on a phase- and intensity-modulated laser. *Opt. Express.* **21**, 29186–94 (2013).

21. D. A. Long *et al.*, Multiheterodyne spectroscopy with optical frequency combs generated from a continuous-wave laser. *Opt. Lett.* **39**, 2688 (2014).

22. F. L. Walls, A. Demarchi, RF Spectrum of a Signal after Frequency Multiplication; Measurement and Comparison with a Simple Calculation. *IEEE Trans. Instrum. Meas.* **24**, 210–217 (1975).

23. D. R. Hjelme, A. R. Mickelson, R. G. Beausoleil, Semiconductor laser stabilization by external optical feedback. *IEEE J. Quantum Electron.* **27**, 352–372 (1991).

24. F. N. Baynes *et al.*, Attosecond timing in optical-to-electrical conversion. *Optica.* **2**, 141 (2015).

25. T. Kobayashi *et al.*, Optical pulse compression using high-frequency electrooptic phase modulation. *IEEE J. Quantum Electron.* **24**, 382–387 (1988).

26. E. Myslivets, B. P. P. Kuo, N. Alic, S. Radic, Generation of wideband frequency combs by continuous-wave seeding of multistage mixers with synthesized dispersion. *Opt Express.* **20**, 3331–3344 (2012).

27. G. P. Agrawal, *Nonlinear Fiber Optics* (Academic Press, 2007).

28. C.-B. Huang, Z. Jiang, D. Leaird, J. Caraquitena, A. Weiner, Spectral line-by-line shaping for optical and microwave arbitrary waveform generations. *Laser Photonics Rev.* **2**, 227 (2008).

29. C. Gohle *et al.*, Carrier envelope phase noise in stabilized amplifier systems. *Opt. Lett.* **30**, 2487 (2005).

30. D. C. Cole, S. B. Papp, S. A. Diddams, Downsampling of optical frequency combs for carrier-envelope offset frequency detection, (available at http://arxiv.org/abs/1310.4134).

31. K. L. Corwin *et al.*, Fundamental Noise Limitations to Supercontinuum Generation in Microstructure Fiber. *Phys. Rev. Lett.* **90** (2003), doi:10.1103/PhysRevLett.90.113904.

32. J. M. Dudley, G. Genty, S. Coen, Supercontinuum generation in photonic crystal fiber. *Rev. Mod. Phys.* **78**, 1135–1184 (2006).

33. D. S. Elliott, R. Roy, S. J. Smith, Extracavity laser band-shape and bandwidth modification. *Phys. Rev. A*. **26**, 12–18 (1982).

34. G. Di Domenico, S. Schilt, P. Thomann, Simple approach to the relation between laser frequency noise and laser line shape. *Appl. Opt.* **49**, 4801 (2010).

35. N. Bucalovic *et al.*, Experimental validation of a simple approximation to determine the linewidth of a laser from its frequency noise spectrum. *Appl. Opt.* **51**, 4582–8 (2012).

36. T. M. Fortier *et al.*, Sub-femtosecond absolute timing jitter with a 10 GHz hybrid photonic-microwave oscillator. *Appl. Phys. Lett.* **100**, 231111 (2012).

37. A. Hati, C. W. Nelson, B. Riddle, D. A. Howe, PM noise of a 40 GHz air-dielectric cavity oscillator (2014) (available at http://arxiv.org/abs/1404.4828).

38. M. Reagor *et al.*, Reaching 10 ms single photon lifetimes for superconducting aluminum cavities. *Appl. Phys. Lett.* **102**, 192604 (2013).